# Antiferromagnetic Topological Insulator MnBi$_2$Te$_4$: Synthesis and Magnetic properties

Hao Li[1,4], Shengsheng Liu[1,5,6,7], Chang Liu[3], Jingsong Zhang[3], Yong Xu[3], Rong Yu[1,5,6,7]*, Yang Wu[2,4]*, Yuegang Zhang[3,4], Shoushan Fan[3,4]

[1]*School of Materials Science and Engineering, Tsinghua University, Beijing, 100084, P. R. China*

[2]*Department of Mechanical Engineering, Tsinghua University, Beijing, 100084, P. R. China*

[3]*State Key Laboratory of Low Dimensional Quantum Physics and Department of Physics, Tsinghua University, Beijing 100084, P. R. China*

[4]*Tsinghua-Foxconn Nanotechnology Research Center, Tsinghua University, Beijing 100084, P. R. China*

[5]*National Center for Electron Microscopy in Beijing, School of Materials Science and Engineering, Tsinghua University, Beijing, 100084, P. R. China*

[6]*Key Laboratory of Advanced Materials of Ministry of Education of China, Tsinghua University, Beijing, 100084, P. R. China*

[7]*State Key Laboratory of New Ceramics and Fine Processing, Tsinghua University, Beijing 100084, P. R. China*

**ABSTRACT :** Recently, MnBi$_2$Te$_4$ has been discovered as the first intrinsic antiferromagnetic topological insulator (AFM TI), and will become a promising material to discover exotic topological quantum phenomena. In this work, we have realized the successful synthesis of high-quality MnBi$_2$Te$_4$ single crystals by solid-state reactions. The as-grown MnBi$_2$Te$_4$ single crystal exhibits a van der Waals layered structure, which is composed of septuple Te-Bi-Te-Mn-Te-Bi-Te sequences as determined by powder X-ray diffraction (PXRD) and high-resolution high-angle annular dark field scanning transmission electron microscopy (HAADF-STEM). The magnetic order below 25 K as a consequence of A-type antiferromagnetic interaction between Mn layers in the MnBi$_2$Te$_4$ crystal suggests the unique interplay between antiferromagnetism and topological quantum states. The transport measurements of MnBi$_2$Te$_4$ single crystals further confirm its magnetic transition. Moreover, the unstable surface of MnBi$_2$Te$_4$, which is found to be easily oxidized in air, deserves attention for onging research on few-layer samples. This study on the first AFM TI of MnBi$_2$Te$_4$ will guide the future research on other potential candidates in the MBi$_x$Te$_y$ family (M = Ni, V, Ti, etc.).

**INTRODUCTION**

During the last decade, tremendous efforts and breakthroughs in topological quantum states have aroused great enthusiasm in pursuing the correlation between topology and symmetry[1,2]. The magnetism and crystal symmetry that affect the time reversal and inversion symmetry, respectively, are undoubtedly critical factors for condensed matters and thus have been deeply investigated in both theoretical and experimental aspects, resulting in numerous novel topological quantum phenomena[1-12]. From the perspective of material science, it is of crucial importance to develop material systems with the coexistence of magnetism and topology that spontaneously breaks the time-reversal symmetry. Most current works tried to introduce magnetism into topological materials extrinsically by doping magnetic impurities[13-18]. However, the extrinsic magnetic defects are difficult to control experimentally, which hinders the progress in subsequent applications, for instance, dissipationless electric transportation in quantum anomalous Hall device and topological quantum computation[1,2]. Hence, synthesizing novel intrinsic magnetic topological materials in which magnetism and topology natively coexist, becomes a key research topic to future applications.

Appealingly, recent first-principle predictions and experiments have demonstrated MnBi$_2$Te$_4$ an intrinsic magnetic topological material, which belongs to antiferromagnetic (AFM) TIs and hosts unparalleled topological physics in both bulk and thin film[19-28]. MnBi$_2$Te$_4$ crystallizes in the tetradymite-type structure with the $R\bar{3}m$ space group and lattice constants a = 4.33 Å, c = 40.91 Å[29-31]. As a member of van der Waals (vdW) layered materials, MnBi$_2$Te$_4$ consists of Te-Bi-Te-Mn-Te-Bi-Te septuple layers (SLs) stacking in the ABC sequence along the c-axis (out-of-plane axis), as depicted in Figure 1a. Therefore, it can be viewed as derived from topological insulator, Bi$_2$Te$_3$, by inserting a Mn-Te layer into the middle of its Te-Bi-Te-Bi-Te quintuple layer (QL). Each Mn$^{2+}$ in high-spin configuration affords 5 $\mu_B$ magnetic moment according to Hund's rule. The intralayer exchange coupling between Mn-Mn is ferromagnetic (FM) along an out-of-plane easy axis, while the interlayer exchange coupling between neighboring SLs is antiferromagnetic (AFM), generating a three-dimensional A-type AFM order[19-23, 31].

Hence, the synthesis of high-quality MnBi$_2$Te$_4$ single crystals with well-defined AFM

order is essential to the research of magnetic topological material, as well as to the potential applications, such as dissipationless transport and low-power electronics. According to previous studies, $MnBi_2Te_4$ can be synthesized through a number of approaches, but lacks of the evidence of high purity crystalline samples[29]. During the preparation of this paper, we noticed that there have been several publications concerning the growth and characterizations of $MnBi_2Te_4$ crystals [20, 30, 31]. However, the synthetic strategy toward high-quality millimeter-sized $MnBi_2Te_4$ single crystals with unambiguous AFM transition was still challenging. More importantly, the oxidation behavior, which is critical to clarify the stability of exfoliated samples, has not been studied yet.

Herein, we report effective synthesis of exclusive $MnBi_2Te_4$ single crystals with a clear and complete antiferromagnetic transition at 25 K. This progress could extend a variety of potential AFM TIs in $MBi_xTe_y$ (M = Ni V, Ti, etc.)[23]. The quality of $MnBi_2Te_4$ single crystals is verified by a number of characterizations in detail. The oxidization behavior of $MnBi_2Te_4$ suggests that the surface of $MnBi_2Te_4$ tends to be oxidized in air while the bulk is stable. Intriguingly, magnetic and transport properties of $MnBi_2Te_4$ single crystals corroborate its AFM order and field-induced magnetic transitions, in good consistence with its prospective intrinsic AFM TI nature.

**RESULTS AND DISCUSSION**

**Crystal Growth and Growth Mechanism.** The challenge in synthesizing ternary intermetallic compounds via solid-state approaches is how to circumvent the uneven distribution of elements in the reaction. To begin with, high-quality $Bi_2Te_3$ and MnTe binary were synthesized as precursors by directly reacting the stoichiometric mixture of high-purity Bi and Te, and Mn and Te, respectively (seen in Method and Figure S1a, b). Afterwards, high-quality single crystal $MnBi_2Te_4$ was grown from a 1:1 mixture of $Bi_2Te_3$ and MnTe. The sample sealed in an evacuated silica ampoule was firstly heated to 973 K, followed by being slowly cooled to 864 K in 7 days and annealed for at least 14 days. After being air-quenched, shiny crystals were selected by cracking the sample chunk. The coarse part of the crystal was carefully cut-off with a scalpel or cleaned off by a scotch tape with the aid of an optical microscope. The growth method can successfully afford $MnBi_2Te_4$ plate-like single crystals up to 3 mm in size with shiny flat surface, indicating its layered structure (cf. inset of Figure 1b).

In fact, attempts by direct reacting of Bi, Mn and Te elements (seen in Method, trial 1) always gave $Bi_2Te_3$ and MnTe as impurities (product 1), which is confirmed by powder X-ray diffraction (PXRD) in Figure S1c. It is reasonable to assume that $MnBi_2Te_4$ forms by the solid-state intercalation of MnTe into $Bi_2Te_3$, which is favored by the long-term annealing. The short-term annealing yields a mixture of $MnBi_2Te_4$, $Bi_2Te_3$ and MnTe (product 2), which is revealed by PXRD in Figure S1d. Furthermore, to gain insights into the reaction mechanism, the high-resolution high-angle annular dark field scanning transmission electron microscopy (HAADF-STEM) was performed on the (0 1 0) crystallographic plane of product 2, and directly shows the mixing of quintuple layers (QLs) from $Bi_2Te_3$ and SLs from $MnBi_2Te_4$ (Figure S2). This demonstrates an uncompleted reaction of $Bi_2Te_3$ and MnTe, which supports aforementioned mechanism. In contrast, for crystal obtained from prolonged annealing, the HAADF-STEM image in Figure 2b clearly resolves its (0 1 0) crystallographic plane with pure SLs from $MnBi_2Te_4$. Therefore, prolonged annealing is essential to guarantee the thorough intercalation of MnTe into $Bi_2Te_3$. The thermal property of $MnBi_2Te_4$ is revealed with differential thermal analysis (DTA). As-grown $MnBi_2Te_4$ crystals display an intense exothermic peak at 872 K on the heating trace, indicating its melting point. Two endothermic peaks at 857 K and 849 K on the cooling run correspond to the crystallization of $MnBi_2Te_4$ and $Bi_2Te_3$ respectively. The lower intensity of endothermic peaks compared to that of the exothermic peak indicates that the crystallization of $MnBi_2Te_4$ is slow and difficult and $MnBi_2Te_4$ might be partly decomposed to $Bi_2Te_3$ above the melting point. The weight loss of $MnBi_2Te_4$ in thermogravimetric (TG) curve (Figure 1d) could result from the volatilization of $Bi_2Te_3$ with a low melting point. As a result, the annealing temperature requires a strict control to avoid the material loss during the crystallization of $MnBi_2Te_4$.

**Characterization.** PXRD was carried out to determine the phase and quality of the as-grown $MnBi_2Te_4$ crystal. Figure 1b shows that the diffraction of raw $MnBi_2Te_4$ crystals only occurs in the sharp and intense peaks that follow the (0 0 l), l=3n, diffraction rule, verifying the rhombohedral symmetry and preferred orientation along the [0 0 1] direction. The PXRD pattern of finely ground $MnBi_2Te_4$ powder in Figure 1b is in good agreement with the calculated PXRD pattern of $MnBi_2Te_4$, demonstrating its $R\bar{3}m$ space group and lattice constants a = 4.33 Å, c = 40.91 Å[29, 30]. No impurity peaks are observed, confirming the high purity of the as-grown $MnBi_2Te_4$ crystal. Besides, Figure 1d shows the Raman spectrum of $MnBi_2Te_4$ crystals with an appreciable blue shift with respect to that of $Bi_2Te_3$, which might result from the stronger in-plane bonding of Mn-Te.

The weak vdW interactions in the layered structure suggest that the top surface of $MnBi_2Te_4$ can be peeled off with scotch tape to expose fresh surface for the following characterization. Figure 2a displays the scanning electron microscope (SEM) image of an exfoliated $MnBi_2Te_4$ crystal with obviously segregated layers, confirming that $MnBi_2Te_4$ is a layered material integrated by van der Waals forces. Energy dispersive X-ray spectroscopy (EDX) was also used to analyze the element distribution of $MnBi_2Te_4$ crystals. As shown in Figure 2c, no peaks other than Mn, Bi, Te, C and Al are observed in the EDX spectrum of $MnBi_2Te_4$. Please note that C and Al is attributed to the conductive tape and sample stage, respectively. The EDX elemental mapping further reveals a uniform element distribution of Bi, Mn and Te (Figure 2d). The element ratio is further confirmed by inductively coupled plasma mass spectrometry (ICP-MS) analysis, which provided that Mn:Bi:Te is 1:2.14:3.96, agreeing with the chemical formula of the title compound. More importantly, the HAADF-STEM image in Figure 2b clearly shows the (0 1 0) crystallographic plane of $MnBi_2Te_4$ single crystal with unambiguously resolved SLs. As the contrast is proportional to the atomic number, the relatively dark atomic layer in the middle of SLs can be assigned to Mn, while two brightest layers are from Bi atoms. Thus, the HAADF-STEM study verifies the stacking sequence of Te-Bi-Te-Mn-Te-Bi-Te in a SL, which agrees well with the structure model. The thickness of a Te-Bi-Te-Mn-Te-Bi-Te SL layer is ~ 1.3 nm according to HAADF-STEM analysis, consistent with one third of the lattice constant of c-axis (Figure 2b).

**Oxidization Behavior and Stability.** We conducted X-ray photoelectron spectroscopy (XPS) measurement to gain insights into oxidation states of $MnBi_2Te_4$. The XPS measurement was firstly carried out on the fresh surface of the $MnBi_2Te_4$ crystal. The XPS survey spectrum verifies the elemental composition (Figure S3a). A low O 1s peak indicates slight surface oxidation of $MnBi_2Te_4$ due to the short exposure to atmosphere during sample transferring.

Furthermore, the oxidation states for each elements can be assigned in the corresponding high-resolution spectra in Figure 3a-c. Two peaks in Bi 4f spectrum at 157.8 eV and 163.1 eV are ascribed to Bi $4f_{7/2}$ and Bi $4f_{5/2}$ of $Bi^{3+}$ in Bi-Te bonds (Figure 3a) [32-34]. As shown in Figure 3b, Te 3d spectrum contains two major peaks at 572.2 eV (Te $3d_{5/2}$) and 582.6 eV (Te $3d_{3/2}$), corresponding to $Te^{2-}$ in Bi-Te bonds and Mn-Te bonds[33, 35]. In addition, two minor peaks at 576.0 eV and 586.3 eV can be observed in Te 3d spectrum, which is attributed to Te-O bonds and indicates slight oxidation of $MnBi_2Te_4$ surface[32]. Figure 3c displays the Mn 2p spectrum, which can be deconvoluted into two sets of four sub peaks (Mn1$^{(')}$, Mn2$^{(')}$, Mn3$^{(')}$, Mn4$^{(')}$). Mn1 and Mn1$'$ peaks are located at 640.1 eV and 651.8 eV, which arises from broken Mn-Te bonds caused by oxidation. The strongest Mn2 and Mn2$'$ peaks at 641.1 and 653.4 eV are contributed by $Mn^{2+}$ in Mn-Te bonds, indicating the major oxidation state of Mn is 2+. Mn3 and Mn3$'$ at 642.5 and 655.5 eV, and Mn4 and Mn4$'$ at 645.8 and 658.6 eV are satellite peaks of Mn1 and Mn1$'$, and Mn2 and Mn2$'$, respectively, which results from the charge transfer between the outer shell of Te and the unfilled 3d shell of Mn in the photoelectron process.[35-38] To reveal the oxidation stability, we exposed $MnBi_2Te_4$ crystals to air over a week, and then investigated the oxidation states with XPS. The survey spectrum in Figure S3b shows an evidently higher O 1s peak, indicating the severer oxidation owing to longer exposure to air. It is also approved by high-resolution XPS spectra. The Bi 4f spectrum in Figure 3d exhibits two strong peaks at 159.2 and 164.4 eV that can be assigned to the Bi-O bonds[32]. Similarly, Te 3d (Figure 3e) and Mn 2p (Figure 3f) spectra of oxidized $MnBi_2Te_4$ surface displays more pronounced oxidation peaks (Te-O) and peaks arising from oxidation (Mn1 and Mn1$'$) than those of fresh $MnBi_2Te_4$ surface. Interestingly, EDX and Raman results of oxidized $MnBi_2Te_4$ in Figure S4 show no obvious difference from those of fresh $MnBi_2Te_4$, implying that the oxidation is a surface behavior of $MnBi_2Te_4$. Therefore, we can conclude that the surface of $MnBi_2Te_4$ is likely to be oxidized and it is critical to keep samples in inert atmosphere especially when handling exfoliated few-layer $MnBi_2Te_4$ samples.

**Magnetic and Transport properties.** Magnetic properties of the as-grown $MnBi_2Te_4$ single crystal were inspected in a superconducting quantum interference device (SQUID) under zero-field-cooled (ZFC) process. Figure 4a and Figure 4b display the temperature dependence of magnetic susceptibility ($\chi$) for the $MnBi_2Te_4$ single crystal in out-of-plane (***H*** // c) and in-plane (***H*** // ab) magnetic fields, respectively. As displayed in Figure 4a, the magnetic susceptibility of $MnBi_2Te_4$ in ***H*** // c increases as temperature decreases and reaches a maximum at 25 K, and then begins to decrease dramatically. This antiferromagnetic transition corresponds to antiferromagnetic order originated from exchange coupling between $Mn^{2+}$ in the neighboring SLs and displays a Neel temperature ($T_N$) of 25 K, which is in good consistence with previous reports[20, 30, 31]. Noticeably, the magnetic susceptibility approaches zero as the $MnBi_2Te_4$ crystal is cooled down to 2 K, implying an ideal antiferromagnetic state with an out-of-plane easy axis. However, the Neel transition tends to become weaker and finally disappears as the applied out-of-plane magnetic field increases, due to the suppression and spin-flop of interlayer antiferromagnetic coupling caused by an external out-of-plane magnetic field[31]. The decrease of $T_N$ with increasing applied magnetic field in Figure S5 also manifests the suppression of magnetic order. The inverse magnetic susceptibility above $T_N$ shows a linear dependence of temperature above $T_N$, which follows the Curie-Weiss (CW) law $\chi = C / (T + \theta_{CW})$. The linear fitting of $\chi^{-1}$ versus $T$ provides a positive CW temperature $\theta_{CW}$ ~ 10 K, and an effective magnetic moment as $\mu_{eff}^c$ ~ 6.2 $\mu_B$ by taking the relationship $C = N_A \mu_{eff}^2 / 3k_B$ into account. This value agrees the spin-only $\mu_{eff}$ of 5.92 $\mu_B$ for high-spin $Mn^{2+}$ ions with $3d^5$ configuration, where $\mu_{eff} = 2\sqrt{J(J+1)} \mu_B$, $J = 5/2$. Comparing with recently published results, it excels the previously reported magnetic moment of 4.04(13) $\mu_B$ per Mn from neutron diffraction[27], and similar to $\mu_{eff} = 5.9(1)$ $\mu_B$ from magnetic measurements[30]. The consistency between measured and calculated $\mu_{eff}$ could originate from the complete intercalation of Mn-Te bilayers into the $Bi_2Te_3$ QLs in this long-term synthesis strategy. We also note that the deviation from the CW law near $T_N$ possibly arises from strong spin fluctuations induced by magnetic phase transition. As to the measurement in ***H*** // ab, the inverse magnetic susceptibility (Figure 4b) also follows the CW law and provides an effective magnetic moment as $\mu_{eff}^{ab}$ ~ 5.4 $\mu_B$, which is close to the spin-only $\mu_{eff}$ ~5.92 $\mu_B$ for a free $Mn^{2+}$ ion. In addition, we can extract a negative Curie-Weiss temperature $\theta_{CW}$ ~ -8 K, implying ferromagnetic interactions among the $Mn^{2+}$, consistent with the predicted A-type AFM feature of $MnBi_2Te_4$. While the magnetic susceptibility decreases dramatically below $T_N$ in ***H*** // c, it only declines slightly below $T_N$ in ***H*** // ab. The obvious difference of the magnetic susceptibility between in-plane and out-of-plane directions suggests an anisotropy of the AFM order. The anisotropic AFM order of $MnBi_2Te_4$ is further revealed in the field dependence of magnetization in Figure 4c and Figure 4d, where the field dependence of magnetization in ***H*** // c and ***H*** // ab is displayed. The magnetization displays a linear dependence and has no hint for saturation for $T > T_N$, indicating the paramagnetic state. In contrast, for ***H*** // c and $T < T_N$, M-H curves increase slowly as the field is increased from zero, and exhibit a sharp increase across a critical field $\mu_0 H_{c1}$ ~ 3.35 T, and turn to be linear for $\mu_0 H > \mu_0 H_{c1}$, suggesting a metamagnetic phase transition into a canted AFM state caused by field-induced suppression of magnetic order[20], in good agreement with $\chi$-$T$ curves. Lastly, the absence of hysteresis in M-H curves is consistent with the AFM state.

Electrical transport measurements were carried out on a $MnBi_2Te_4$ single crystal with a thickness of ~ 10 μm. Figure 5a displays the metallic temperature dependence of longitudinal resistivity $\rho_{xx}$ from 1.5 K to room temperature. A sharp transition at ~ 25 K corresponding to the AFM transition is consistent with $T_N$ from the magnetic property measurement. The magneto-resistivity (MR) curves measured in ***H*** // c at different temperatures are shown in Figure 5b. As the magnetic field is increased from zero, a sharp decrease of $\rho_{xx}$ takes place over the critical magnetic field ($H_{c1}$) below $T_N$, which indicates the metamagnetic phase transition consistent with forenamed magnetic properties. When further increasing the magnetic field, a kink followed by a gradual decrease in $\rho_{xx}$ emerges, suggesting that the fully polarized spin in $Mn^{2+}$ occurs at a higher critical magnetic field ($H_{c2}$). The similar electric transport behavior with both critical magnetic fields $H_{c1}$, $H_{c2}$ were also observed and reported in recent publications[20, 24, 27]. Figure 5c shows the field dependence of Hall resistivity $\rho_{yx}$ at the same temperatures. The negative slopes indicate that electron is the dominating charge carrier, possibly caused by defects such as Te deficiency or antisite disordering[30]. From the linear region of Hall curves ($H < H_{c1}$), the electron density is extracted to be $2.3 \times 10^{20}$ cm$^{-3}$, and the mobility is calculated to be ~ 630 cm$^2 \cdot V^{-1} \cdot s^{-1}$ at 1.5 K. Similarly, Hall curves exhibit a sharp change in the slope across $H_{c1}$ and a twist across $H_{c2}$ below $T_N$, which conforms to the sharp decrease and kink at $H_{c1}$ and $H_{c2}$ in $\rho_{xx}$ (Figure 5b), respectively.

**CONCLUSIONS**

In summary, we have developed a facile strategy to synthesize high-quality MnBi$_2$Te$_4$ single crystal by prolonged annealing across a narrow temperature window. The crystal structure and quality of MnBi$_2$Te$_4$, especially its septuple Te-Bi-Te-Mn-Te-Bi-Te sequences and oxidation behavior, have been revealed via various characterizations. The as-grown MnBi$_2$Te$_4$ single crystal exhibits an evident AFM transition at 25 K and field-induced magnetic transitions in both magnetic and transport properties, manifesting its intrinsic AFM nature. This study on the first AFM TI of MnBi$_2$Te$_4$ could open an avenue on novel quantum states and topological phenomena and further inspire research on other potential AFM TIs of MBi$_x$Te$_y$ (M = Cr, V, Ti, etc.)[23].

**EXPERIMENTAL SECTION**

**Materials.** Bi (99.99%, Adamas), Mn (99.95%, Alfa Aesar), Te (99.999%, Aladdin) used as received.

**Method.** (a) Synthesis of MnTe. Polycrystalline MnTe was synthesized by directly heating the stoichiometric mixture of high-purity Mn and Te at 1273 K in a vacuum-sealed silica ampoule for 3 days. (b) Synthesis of Bi$_2$Te$_3$. Bi$_2$Te$_3$ crystal was synthesized by a direct solid-state reaction of the stoichiometric mixture of high-purity Bi and Te. The reaction mixture was vacuum-sealed in a silica ampoule, and heated to 1083 K and held for 24 h. After being slowly cooled to 833 K at a rate of 0.1 K·min$^{-1}$, the ampoule was quenched in air to obtain Bi$_2$Te$_3$ crystal. (c) Trail 1. The product 1 was synthesized by direct heating of the stoichiometric mixture of high-purity Mn, Bi and Te at 973 K in a vacuum-sealed silica ampoule for 24 h.

**Characterization.** The powder XRD patterns were collected on a PANalytical Empyrean X-ray diffractometer using Cu Kα radiation. Raman spectra were collected on a Horiba Jobin Yvon LabRam-HR/VV Spectrometer with a 514 nm laser and an 1800 mm$^{-1}$ grating. DTA and TG analysis (TG) were carried out on a NETZSCH STA449 F3 simultaneous thermal analyzer under N$_2$ atmosphere, and the sample was enclosed in an Al$_2$O$_3$ crucible. The heating rate was 10 K min$^{-1}$ from room temperature to 953 K, and then cooled to 573 K at a rate of 3 K min$^{-1}$. The SEM images and EDX spectra were collected on a FEI NOVA SEM450 scanning electron microscope. The ICP-MS analysis was carried out on a Thermo Fisher iCP QC inductively coupled plasma mass spectrometer. The HAADF-STEM images were collected on a Titan Cubed Themis G2 300 Double Aberration-Corrected Transmission Electron Microscope, while the sample for cross-sectional observation was prepared via a Zeiss Auriga focused ion beam (FIB) instrument. The XPS spectra were collected on a ULVAC PHI Quantera II X-ray Photoelectron Spectrometer.

**Magnetic Property Measurement.** Magnetic Property Measurement were carried out on a Quantum Design Superconducting Quantum Interference Device using a vibrating sample magnetometer.

**Transport measurement.** Electrical transport measurements were carried out in an cryostat (Oxford Instruments) with a base temperature of ~ 1.5 K and a magnetic field up to 9 T. The longitudinal and Hall voltages were detected simultaneously by using Stanford Research Instrument SR830 lock-in amplifiers with an AC current generated with a Keithley 6221 current source.

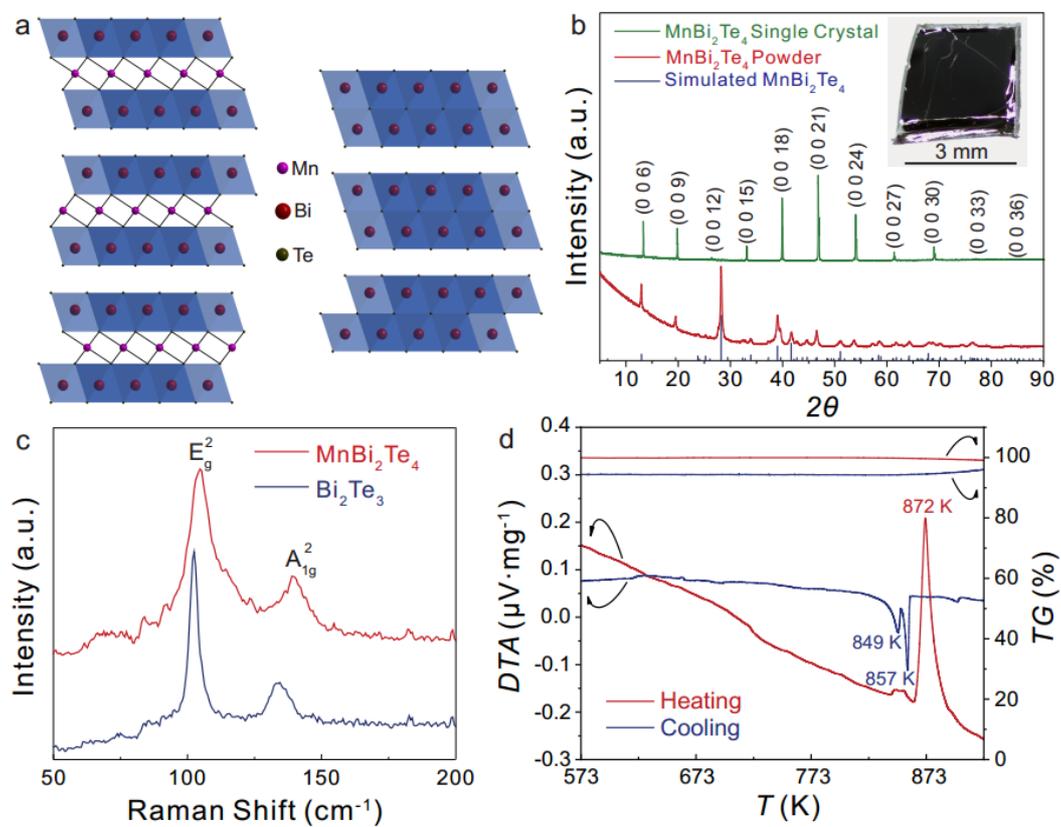

**Figure 1.** (a) Schematic of crystal structure of the (1 1 0) crystallographic plane of $MnBi_2Te_4$ (left) and $Bi_2Te_3$ (right). (b) PXRD pattern of $MnBi_2Te_4$ single crystal (top) and powder (down). Inset: an optical image of the as-grown $MnBi_2Te_4$ single crystal. (c) Raman spectra of $MnBi_2Te_4$ and $Bi_2Te_3$. (d) Differential thermal analysis (DTA) and thermogravimetric (TG) curves for crystalline $MnBi_2Te_4$.

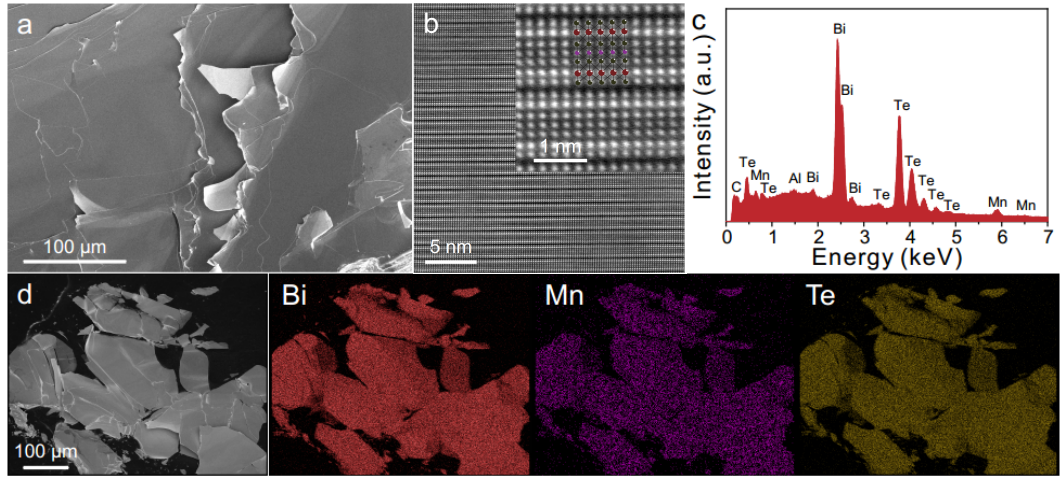

**Figure 2.** (a) SEM image of exfoliated MnBi$_2$Te$_4$ single crystal. (b) HAADF-STEM image for the (0 1 0) crystallographic plane of MnBi$_2$Te$_4$, in which SLs can be clearly resolved. Inset: enlarged HAADF-STEM image superimposed with the schematic structure of the (0 1 0) crystallographic plane of MnBi$_2$Te$_4$. (c) EDX spectrum of MnBi$_2$Te$_4$ in (a). (d) SEM image and the corresponding EDX elemental mapping (Bi, Mn and Te).

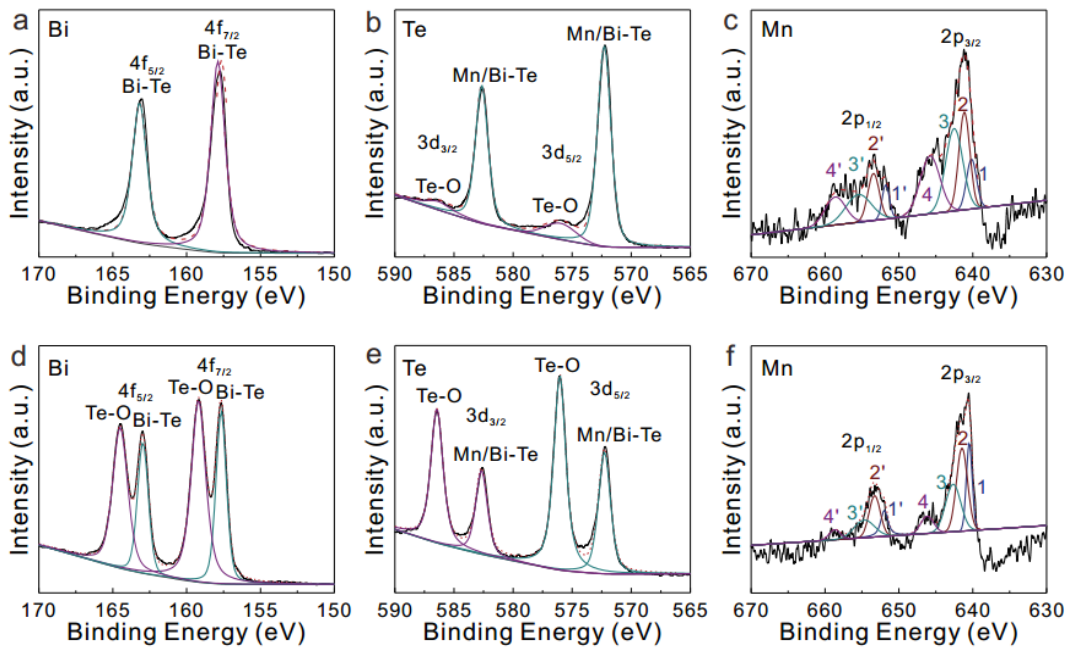

**Figure 3.** High-resolution XPS spectra of fresh (a, b, c) and oxidized (d, e, f) surface of MnBi$_2$Te$_4$, respectively. (a, d) Bi 4f, (b, e) Te 3d and (c, f) Mn 2p spectra.

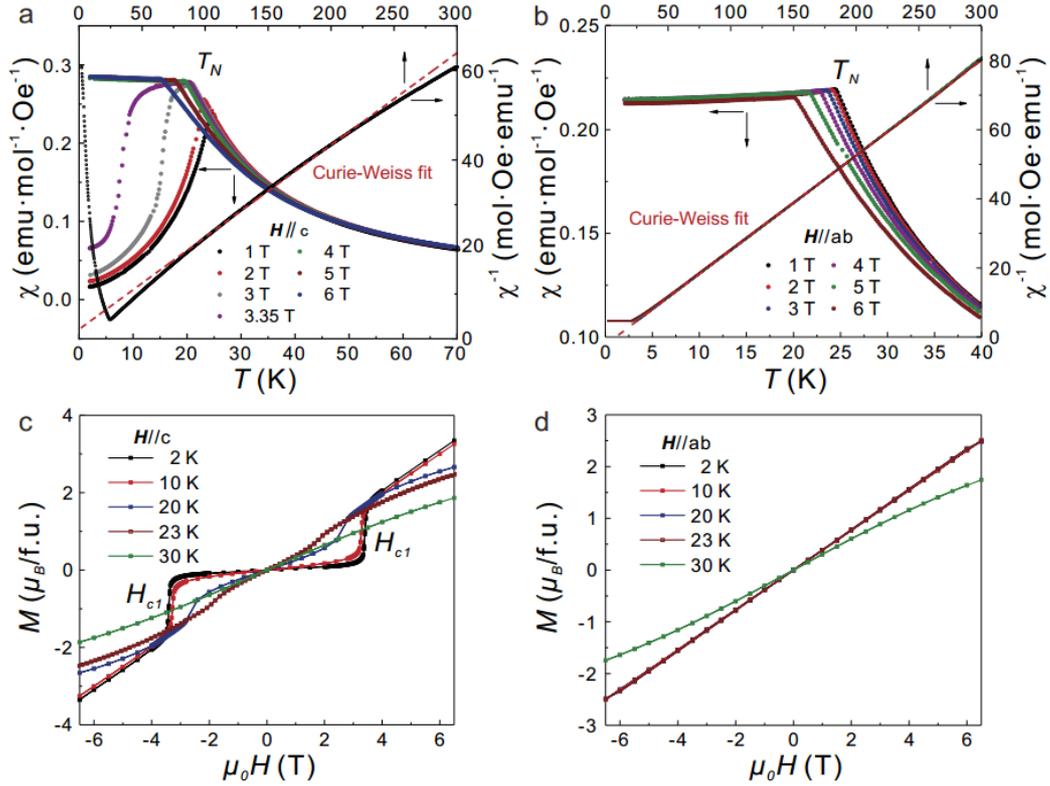

**Figure 4. Magnetic properties of MnBi$_2$Te$_4$ single crystal.** (a, b) Temperature dependence of magnetic susceptibility for the MnBi$_2$Te$_4$ single crystal measured under ZFC process and inverse magnetic susceptibility as a function of temperature (top x axis and right y axis). The symbol and dash line represent the experimental data and Curie-Weiss fit. Magnetic field is out-of-plane (a) and in-plane (b), respectively. (c, d) Field dependence of magnetization for the MnBi$_2$Te$_4$ crystal measured at different temperatures. Magnetic field is out-of-plane (c) and in-plane (d), respectively.

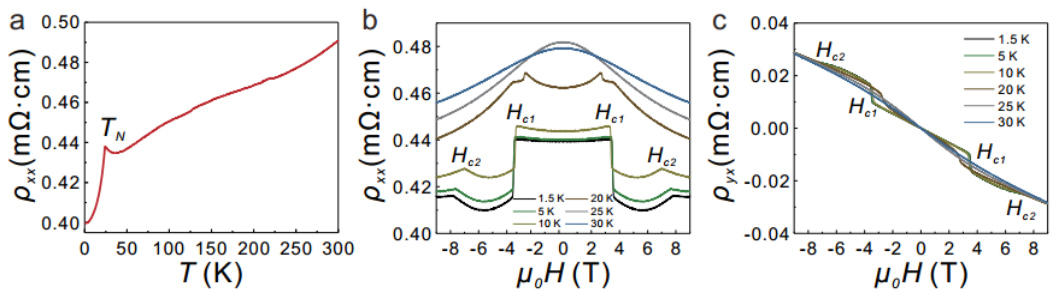

**Figure 5. Transport properties of MnBi$_2$Te$_4$ single crystal.** (a) Temperature dependence of longitudinal resistivity $\rho_{xx}$ measured from 1.6 K to room temperature. The sharp AFM transition is revealed near $T_N$ = 25 K. (b) MR curves in an out-of-plane magnetic field at varied

temperatures. The lower and upper critical fields ($H_{c1}$, $H_{c2}$) are labeled and the temperatures for MR curves are shown by different color from 30 K to 1.5 K. (c) Magnetic field dependence of Hall resistivity traces measured at the same temperatures.

## ASSOCIATED CONTENT

**Supporting Information**

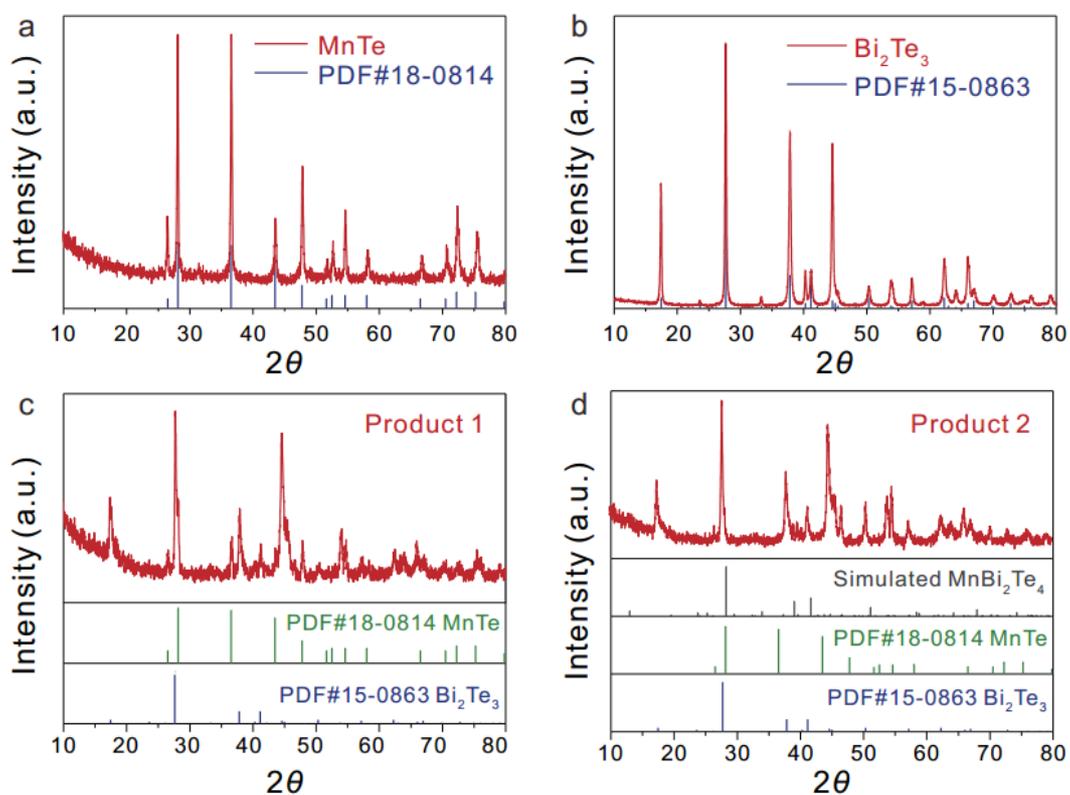

**Figure S1.** Experimental and reference PXRD patterns of (a) MnTe, (b) $Bi_2Te_3$ and (c) Product 1. (d) Product 2.

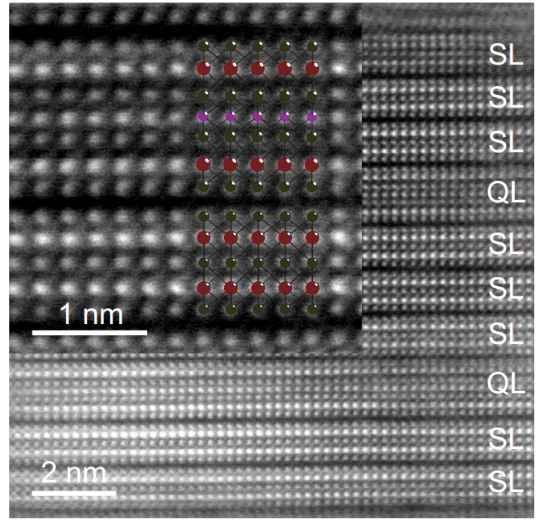

**Figure S2.** HAADF-STEM image for the (0 1 0) crystallographic plane of Product 2. Inset: enlarged HAADF-STEM image superimposed with the schematic structure of the (0 1 0) crystallographic plane of $MnBi_2Te_4$ and $Bi_2Te_3$.

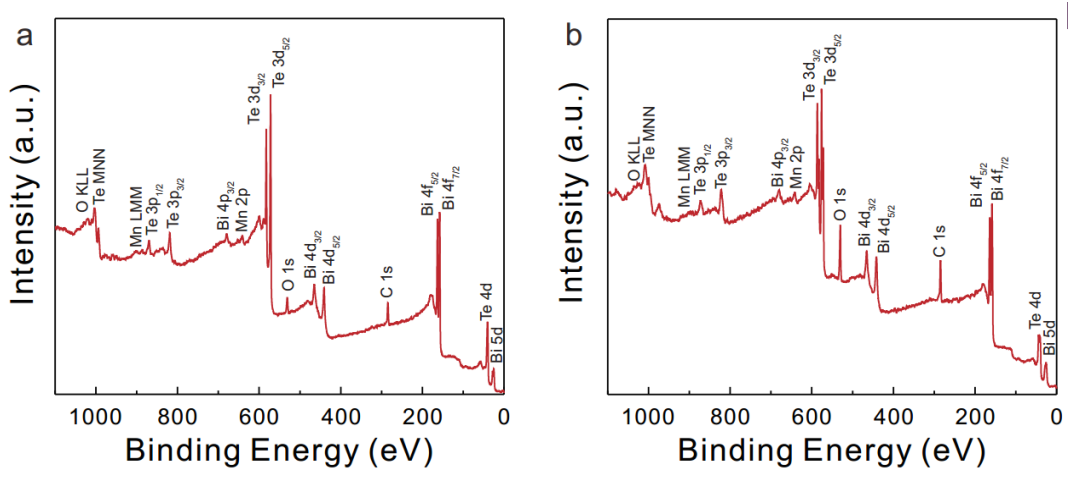

**Figure S3.** Survey XPS spectra of (a) fresh and (b) oxidized $MnBi_2Te_4$ surfaces.

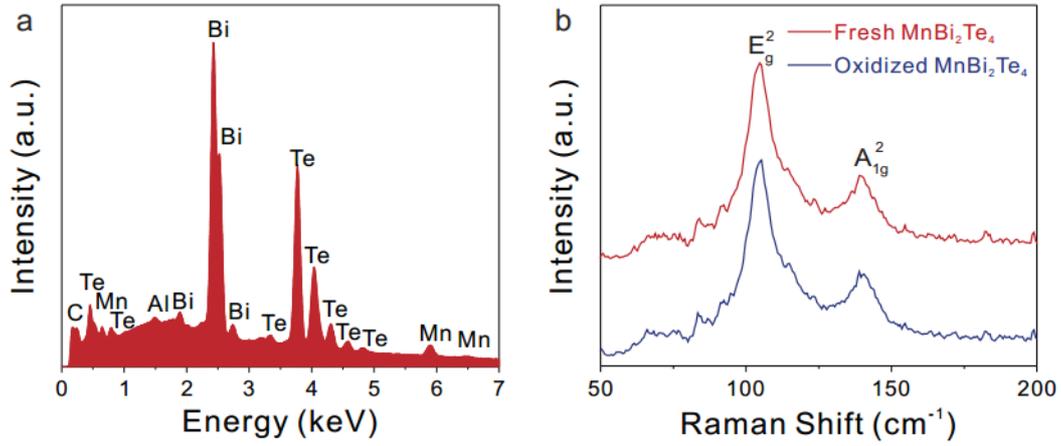

**Figure S4.** (a) EDX spectrum of oxidized $MnBi_2Te_4$. (b) Raman spectrum of oxidized and fresh $MnBi_2Te_4$.

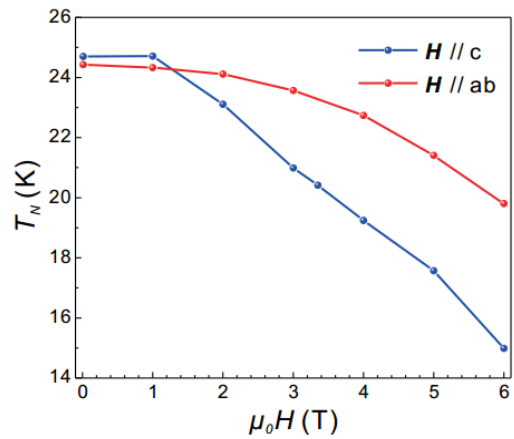

**Figure S5.** Suppression of $T_N$ with increasing applied magnetic fields.

## AUTHOR INFORMATION

**Corresponding Author**

* E-mails: wuyangthu@tsinghua.edu.cn, ryu@tsinghua.edu.cn.

## ACKNOWLEDGMENTS

This work was supported by the Basic Science Center Project of NSFC (Grant No. 51788104), the Ministry of Science and Technology of China (Grants No. 2018YFA0307100, and No. 2018YFA0305603), and the National Natural Science Foundation of China (Grant No. 11674188, U1832218).